\documentclass[doublecol]{epl2}

\title{Many-body properties of quasi-one dimensional Boson gas across  a narrow CIR}

\author{Ran Qi\inst{1} \and Xiwen Guan\inst{2,3}}
\shortauthor{Ran Qi \etal}

\institute{
  \inst{1} Institute for Advanced Study, Tsinghua University, Beijing, 100084, China\\
  \inst{2} State Key Laboratory of Magnetic Resonance and Atomic and Molecular Physics, Wuhan Institute of Physics and Mathematics, Chinese Academy of Sciences, Wuhan 430071, China\\
  \inst{3} Department of Theoretical Physics, Research School of Physics and Engineering, the Australian National University, Canberra ACT 0200, Australia
}

\abstract{We study strong interaction effects  in a  one-dimensional (1D)  Boson gas across a narrow confinement induced resonance (CIR).
  In contrast to the zero range potential, the 1D two-body interaction in the narrow CIR can be written as a polynomial of  derivative $\delta$-function interaction on many-body level.   Using the asymptotic  Bethe ansatz,  we find that the low energy physics of this  many-body problem  is  described by  the   Tomonaga-Luttinger liquid  where the  Luttinger parameters are essentially  modified by an effective  finite  range parameter  $v$.  This parameter drastically alters quantum criticality  and universal thermodynamics of the  gas. In particular, it  drives   the  Tonks-Girardeau (TG)  gas   from  non-mutual  Fermi statistics  to  mutual statistics or  to a more exclusive super TG  gas.  This novel feature is further discussed in terms of the breathing mode which is experimentally measurable.}

\begin{document}

\maketitle

\section{Introduction}
Over the past few decades,  experiments on  ultracold bosonic and fermionic atoms confined to one dimensional (1D) geometry \cite{Tonk,Kinoshita,Liao} have provided a better understanding of significant quantum statistical and strong interaction effects in quantum many-body systems. The experimental measurements   to date are seen  to be in good agreement with results obtained from  exactly solved models \cite{Lieb,Lieb2,Yang:1967}. In particular, Haller et al. \cite{superTonk} made an experimental breakthrough in 2009 by realizing a stable highly excited gas-like phase - called the super Tonks-Girardeau (TG) gas -  in the strongly attractive regime of bosonic Cesium atoms. Such a gas was predicted earlier by Astrakharchik {\em et al.} \cite{Astra}  on a gas of hard rods and by Batchelor {\em et al.} \cite{Batchelor} from the Bethe ansatz (BA) integrable Bose gas with an attractive interaction.

So far most  quasi-1D confined systems have been experimentally realized  by the wide Feshbach resonance with combination of confinement induced resonance (CIR).  At low energies, the quasi-1D atom-atom interaction can be written  as  a $\delta$-function potential $V(x)=g_{1d}\delta(x)$  with an effective  interacting strength $g_{1d}=c_0/m$. Where $c_0=2[a_{\bot}(a_{\bot}/a_s-C)]^{-1}$ is determined  by  the 3D scattering length $a_s$.  Here $a_{\bot}$ is the confining length and $C=1.0326$ is a constant  \cite{CIR}.  Across a CIR ($a_s\sim a_{\bot}$), $g_{1d}$ can be tuned from $+\infty$ to $-\infty$. Consequently, the quantum dynamics,  correlations  and thermodynamics  of  few quasi-1D systems have been experimentally elucidated.

However, the resonance width has an essential effect  which affects  interacting properties  on many-body level.   It has been recently pointed out \cite{Ho} that  a 3D narrow Feshbach resonance may produce strong interaction effects due to the resonance structure of its phase shift.  In the  narrow resonance,  many-body  properties are not only determined by the scattering length $a_s$ but also by an effective range that introduces  a strong energy-dependent  scattering amplitude of two colliding atoms \cite{Gurarie,xiaoling}.  In this paper, we  find  that the 1D two-body interaction across a narrow CIR leads to significant  finite range dependent interaction effects in many-body properties of the model, including universal  Tomonnaga-Luttinger  liquid (TLL) physics, thermodynamics, quantum criticality, super TG phase and collective modes, etc.

\section{Two-body scattering property and quasi-1D  confined system}
In  narrow resonances, the two-body scattering physics is well described by an energy dependent scattering length $a_{s}^{-1}(E)=a_s^{-1}(0)-mr_0E$, where $a_s(0)$ is zero energy scattering length and $E$ is scattering energy. Here $r_0$ denotes the effective range and it could be large and negative in the narrow Feshbach resonances.  By solving the two-body problem under a tight harmonic confinement in two radial directions, one obtains the effective 1D scattering amplitude \cite{xiaoling}
\revision{
\begin{eqnarray}
f_k=-\frac{1}{1-i2k/c(k)}\label{fk}
\end{eqnarray}
}
where $c(k)=(c_0^{-1}+4vk^2)^{-1}$, $c_0^{-1}=a_{\bot}[a_{\bot}/a_s(\omega_{\bot})-C]/2$ and $v=-a_{\bot}^2r_0/8$. The confining length $a_{\bot}=\sqrt{1/(m\omega_{\bot})}$  is determined by the trapping frequency $\omega_{\bot}$.  The CIR corresponds to the limit $c_0\rightarrow \infty$. The only difference between narrow and wide CIRs is that  the interacting strength $c(k)$, instead of being a constant, depends on the relative momentum of two collision  atoms.  Under this circumstance,  the energy dependent scattering amplitude is characterized by the parameter $v$ which has a dimension of $L^3$. For an infinite wide resonance, we have $r_0\rightarrow 0$ then $v=0,~c(k)\equiv c_0$ and  the scattering amplitude  $f_k$ becomes identical to the zero range one  \cite{CIR}.

Remarkably, we prove that for $v\neq 0$ the two-body scattering amplitude leads to  the following Hamiltonian  with a polynomial of derivative $\delta$-function  interaction on many-body level
\begin{eqnarray}
&&\hat{H}=-\frac{1}{2m}\sum_{i=1}^N\frac{\partial^2}{\partial x_i^2}\label{model}\\
&&+\frac{c_0}{m}\sum_{i<j}\delta(x_i-x_j)\Big[1\!+\!\sum_{\ell=1}^{\infty}(c_0v)^{\ell}\Big(\frac{\partial}{\partial x_i}-\frac{\partial}{\partial x_j}\Big)^{2\ell}\Big].\nonumber
\end{eqnarray}
Where $m$ is atom mass and $N$ is the total particle number.
To justify equation (\ref{model}) as a proper description of our narrow CIR system, we will show that Hamiltonian (\ref{model}) leads to the correct scattering amplitude given in equation (\ref{fk}) by solving the two-body scattering wave function. In a two-body problem, Hamiltonian (\ref{model}) becomes
\begin{eqnarray}
&&\hat{H}=-\frac{1}{2m}\frac{\partial^2}{\partial x_1^2}-\frac{1}{2m}\frac{\partial^2}{\partial x_2^2}\\
&&+\frac{c_0}{m}\delta(x_1-x_2)\Big[1\!+\!\sum_{\ell=1}^{\infty}(c_0v)^{\ell}\Big(\frac{\partial}{\partial x_1}-\frac{\partial}{\partial x_2}\Big)^{2\ell}\Big].\nonumber
\end{eqnarray}
Since the interaction is still contact in nature, we can choose the following form of two-body wave function in the region $x_1<x_2$
\begin{eqnarray}
\psi(x_1,x_2)=e^{ik_1x_1+ik_2x_2}+A_{k_1,k_2}e^{ik_2x_1+ik_1x_2}.
\end{eqnarray}
The coefficient $A_{k_1,k_2}$ is determined by the boundary condition \cite{Lieb}
\begin{eqnarray}
\frac{1}{m}\Big(\frac{\partial}{\partial x_2}-\frac{\partial}{\partial x_1} \Big)\psi|_{x_1=x_2}=\frac{c\big(\frac{k_1-k_2}{2} \big)}{m}\psi|_{x_1=x_2}
\end{eqnarray}
where $c(k)$ was defined below equation (\ref{fk}) and we have used the relation $(1+x)^{-1}=1+\sum_{\ell=1}^\infty(-x)^{\ell}$. This boundary condition gives:
\begin{eqnarray}
 A_{k_1,k_2}=-\frac{c\big(\frac{k_1-k_2}{2} \big)+i(k_1-k_2)}{c\big(\frac{k_1-k_2}{2} \big)-i(k_1-k_2)}.
\end{eqnarray}
Finally, the two-body scattering amplitude is given as
\begin{eqnarray}
f(k)=\frac{1}{2}(A_{k,-k}-1)=-\frac{1}{1-i2k/c(k)}
\end{eqnarray}
which is exactly the same as equation (\ref{fk}). As a result, we have proved that Hamiltonian (\ref{model}) indeed reproduces the correct two-body scattering amplitude of a narrow CIR thus justifies the using of this model.

Below, we will try to solve the full many-body Hamiltonian (\ref{model}) analytically. As mentioned above, the interactions in model (\ref{model}) is still contact in nature. Therefore, in the knowledge of asymptotic BA \cite{Sutherland},
for  the domain $x_1<\cdots<x_N$, the wave function of the model (\ref{model}) can be written as  as  super positions of plane waves  \cite{Gutkin,xiwen}, i.e.
\begin{eqnarray}
\Psi(x_1,x_2,\cdots,x_N)=\sum_P A_P e^{i\sum_j k_{P_j}x_j},
\end{eqnarray}
where $P$ runs over all $N!$ permutations, the coefficients $A_P=(-1)^P\prod_{j<\ell}^N\frac{1+i(k_{P_{\ell}}-k_{P_j})[c_0^{-1}+v(k_{P_{\ell}}-k_{P_j})^2]}{\sqrt{1+(k_{P_{\ell}}-k_{P_j})^2[c_0^{-1}+v(k_{P_{\ell}}-k_{P_j})^2]^2}}$. The wave functions in other domains can be obtained from the interchange symmetry of bosons.
The eigenvalue of the model (\ref{model})   is given by $E=\sum_{j=1}^Nk^2_j$  where  the  wave numbers $\{k_j \}$ satisfy  the following transcendental equations
\begin{eqnarray}
k_jL&=&2\pi I_j-\sum_{\ell=1}^N \theta(k_j-k_{\ell}).\label{kj2}
\end{eqnarray}
Here we denoted $\theta(p)=2\arctan[p(c_0^{-1}+vp^2)]$.  $I_j $ is an integer for odd $N$ and a half odd integer for even $N$.   In this work, we only consider the repulsive scattering branch corresponding to the solutions with all $k_j$ being real.

\section{Ground state  and low energy physics}
\begin{figure}[t]
\includegraphics[height=1.5in, width=3.3in]
{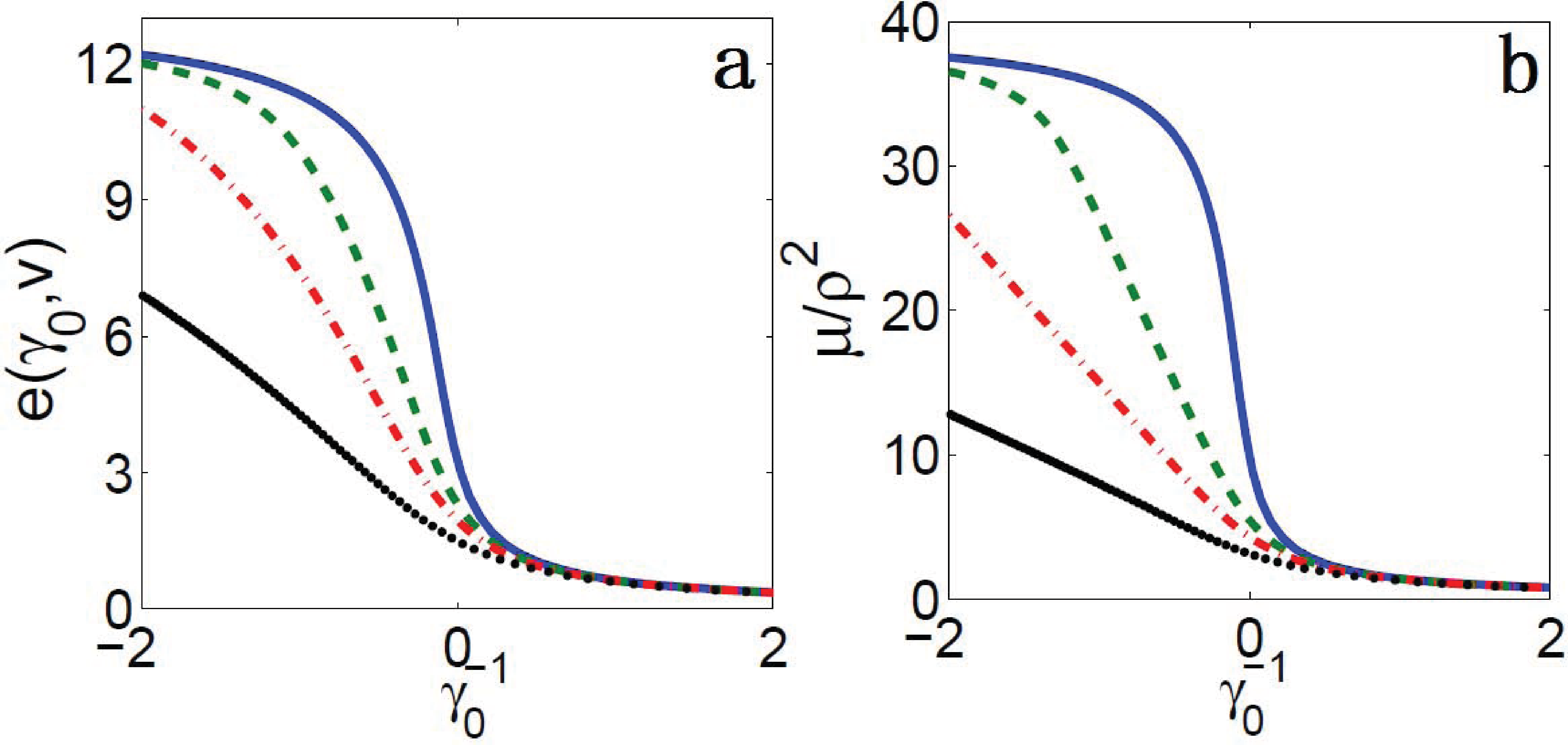}
\caption{The energy density and chemical potential  vs  zero energy interacting strength $r_0^{-1}$ for the gas-like phase. The curves stand for the values of  $\tilde{v}=0,~0.01,~0.02,~0.05$ (from top to bottom).   \label{GSenergy}}
\end{figure}

In the thermodynamic limit $N,~L\rightarrow \infty$ with finite particle density $n=N/L$, we can define a  quasi momentum distribution $\rho_k$. At the  ground state, $\rho_k$ is non-zero only in a finite interval $[-Q,\, Q]$ where $Q$  is the ``Fermi momentum" of the 1D  interacting bosons.  From equation (\ref{kj2}) we have $\rho_k$ satisfying
\begin{eqnarray}
\rho_k&=&\frac{1}{2\pi}\Big[1+\int_{-Q}^{Q} K(k-q)\rho_q dq\Big]\label{rhokground}
\end{eqnarray}
where $K(p)=\frac{\partial\theta(p)}{\partial p}=\frac{2(c_0^{-1}+3vp^2)}{1+p^2(c_0^{-1}+vp^2)^2}$ and the particle density $n=\int_{-\infty}^{+\infty}\rho_kdk$.
 Thus the  ground state energy $E_0$ and chemical potential $\mu$ are  given by
\begin{eqnarray}
E_0&=&L\int_{-Q}^Q k^2\rho_k dk=Ln^3e(\gamma_0,\tilde{v})\\
\mu&=&\frac{\partial E_0}{\partial N}=n^2\Big(3e-\gamma_0\frac{\partial e}{\partial \gamma_0}+3\tilde{v}\frac{\partial e}{\partial \tilde{v}} \Big)\label{mu}
\end{eqnarray}
where $\gamma_0=c_0/n,~\tilde{v}=vn^3$ are dimensionless counterpart of $c_0$ and $v$ ($v<c_0$).  By analysis of the BA equation (\ref{rhokground}), we see that the finite range effect  does not play a role in weak coupling regime. However, for a strong interacting regime, $v$ drastically changes the ground  energy. For  $\gamma_0\to \infty$, the kernel function becomes $K(p)=\frac{6vp^2}{1+v^2p^6} $. This means that the effective finite range $v$ can drive the the TG gas with non-mutual Fermi statistics \cite{Haldane} into the gas-like  phase with  mutual statistics  either less or more exclusive than the free Fermion statistics, i.e. super TG gas. Fig. \ref{GSenergy}  shows  strong interaction effects in $e(\gamma_0,\tilde{v})$ and $\mu$ at different values of  $\tilde{v}$ \cite{Liming}.
One sees that the energy and chemical potential  are significantly reduced  as $\tilde{v}$  increases  in  the TG and super-TG gas-like phases (in accordance with the wide CIR, we here refer  the TG and super-TG regimes   for  $\gamma_0\to \infty$ and $\gamma_0\to -\infty $). It indicates  that the zero range repulsive interaction strength is reduced  by the positive $v$ at the collision.

The elementary excitations of  the model (\ref{model}) are characterized by the quantum numbers $\left\{ I_j \right\}$ for the BA equations (\ref{kj2}), including moving a particle close to the right or left Fermi points out sides the Fermi sea  and moving a particle from the left Fermi point  to the right \cite{Lieb}.  Here we find  that the elementary  spectra are  phonon like  in the long wave length limit and have the same sound velocity. The sound velocity $\nu_c$ can  be obtained from the BA equations  (\ref{kj2}) by
\begin{eqnarray}
\nu_c=\lim_{p\rightarrow 0}\frac{\omega_p}{p}=\frac{Q}{m}\frac{1+\frac{1}{Q}\int_{-Q}^Qkf_kdk}{1+\int_{-Q}^Qf_kdk}, \label{vc1}
\end{eqnarray}
where $f_k$ satisfies the equation $f_k=\big[\int_{-Q}^Q\!K(k\!-\!q)f_qdq\!+\!K(Q\!-\!k)\big]/(2\pi)$ and $\omega_p$ refers to the energy of the  elementary  particle-hole excitations.

\begin{figure}[t]
\includegraphics[height=1.5in, width=3.2in]
{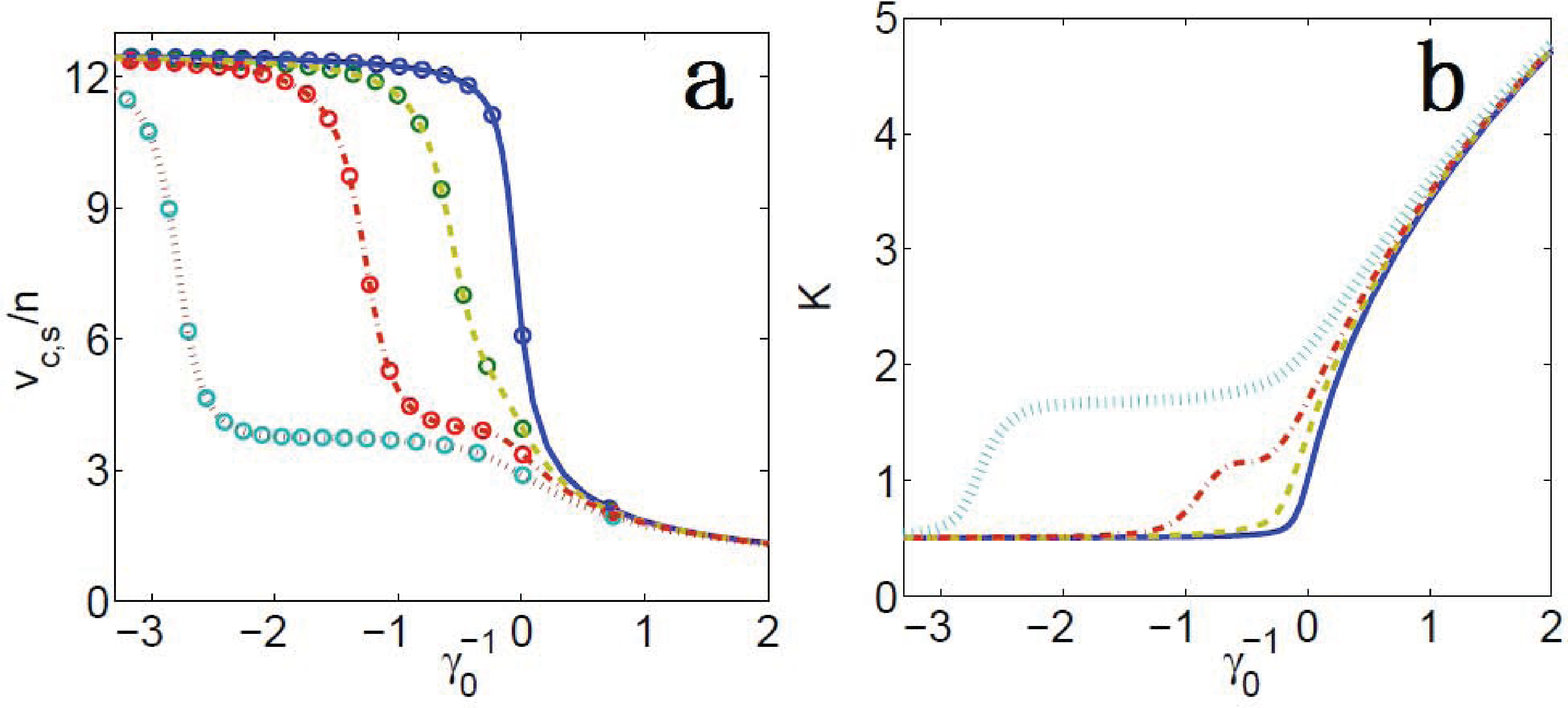}
\caption{ Sound velocity and Luttinger parameter $K$ vs $r_0^{-1}$ for  the values of $\tilde{v}=0,~0.005,~0.01,~0.02$ (solid, dashed, dash-dotted and dotted lines).  In  figure $(a)$, all  lines  stand for  result of $\nu_s=\sqrt{\nu_J\nu_N}$ whereas  the open circles present the  $\nu_c$ from the elementary excitation spectrums (\ref{vc1}).
   \label{TLLpic}}
\end{figure}

On the other hand,  the low energy physics of a wide class of 1D interacting models can be captured by the TLL described by the following effective Hamiltonian \cite{Bosonization1,Bosonization2}:
\begin{eqnarray}
H_{eff}=\frac{1}{2\pi}\int_0^Ldx [\nu_J(\partial_x\phi(x))^2+\nu_N(\partial_x\theta(x))^2]\label{TLL}
\end{eqnarray}
where $\phi(x)$ and $\theta(x)$ correspond to phase and density fluctuations. For the Galilean invariance, $\nu_J=\nu_sK=n\pi/m$ and $\nu_N=\nu_s/K$ are phase and density stiffness respectively. Here $K$ is the Luttinger parameter which determines  the critical exponents of correlation functions. The Hamiltonian (\ref{TLL}) also has  a linear spectrum $\omega_p=\nu_s p$ with $\nu_s=\sqrt{\nu_J\nu_N}$. The velocity $\nu_N=\frac{1}{\pi}\frac{\partial\mu}{\partial n} $ can be obtained from our exact solution (\ref{mu}).
In Fig. \ref{TLLpic}, we verify  from $\nu_s=\nu_c$ that the effective model (\ref{TLL}) indeed captures the collective low energy physics of model (\ref{model}).
This conclusion will be  further identified from the universal leading order finite temperature corrections to the free energy.
 The figure $b$ in  Fig. \ref{TLLpic} shows   influences of the effective  finite range $v$ on the  Luttinger parameter $K=\nu_s/\nu_N$.

\section{Universal thermodynamics}
At finite temperatures, the true physical states can be determined from the minimization of the Gibbs free energy. Following Yang-Yang method  \cite{Yang}, the  equation of state can be obtained from  the pressure
\begin{eqnarray}
P&\!=\!&\frac{T}{2\pi}\int_{-\infty}^{+\infty}\ln(1+e^{-\epsilon_k/T})dk\label{P1}
\end{eqnarray}
where the ``dressed energy" $\epsilon_k$ satisfies the so-called thermodynamic BA (TBA)  equation
\begin{eqnarray}
\epsilon_k\!=\!k^2\!-\!\mu\!-\!\frac{T}{2\pi}\int_{-\infty}^{+\infty}\! K(k-q)\ln(1+e^{-\epsilon_q/T})dq.\label{TBA}
\end{eqnarray}
This equation provides an analytical way to access full thermodynamics of the many-body system  (\ref{model}).

At high temperatures,  the  fugacity $z=e^{\mu/T}\ll 1$, thus we can perform high temperature  expansions  from the pressure  (\ref{P1}), namely,
\begin{eqnarray}
P&=&P^{(0)}+\frac{T^{3/2}}{\sqrt{2\pi}}\Delta b_2 z^2+O(z^3)\label{phighT}\\
\Delta b_2&=&-\frac{1}{2}+\frac{1}{\pi}\int_0^{\infty}\frac{d\delta(k)}{d k}e^{-2k^2/T}dk\label{b2}
\end{eqnarray}
where $\delta(k)=\arctan[2k/c(k)]$ is the two-body scattering phase shift.  The noninteracting part is given by
$P^{(0)}=-\frac{T}{2\pi}\int_{-\infty}^{+\infty}\ln(1-ze^{-k^2/T})dk$.  The second viral coefficient
 $\Delta b_2$ in (\ref{b2}) fully agrees with the  result  obtained  from  the  pure two-body physics \cite{xiaoling}. In fact, the second term in (\ref{phighT}) contains subtle  corrections from quantum statistical and  dynamical  interactions as well as  the finite range  effect  to the pressure of  the Boltzmann gas.

\begin{figure}[t]
\includegraphics[height=1.5in, width=3.4in]
{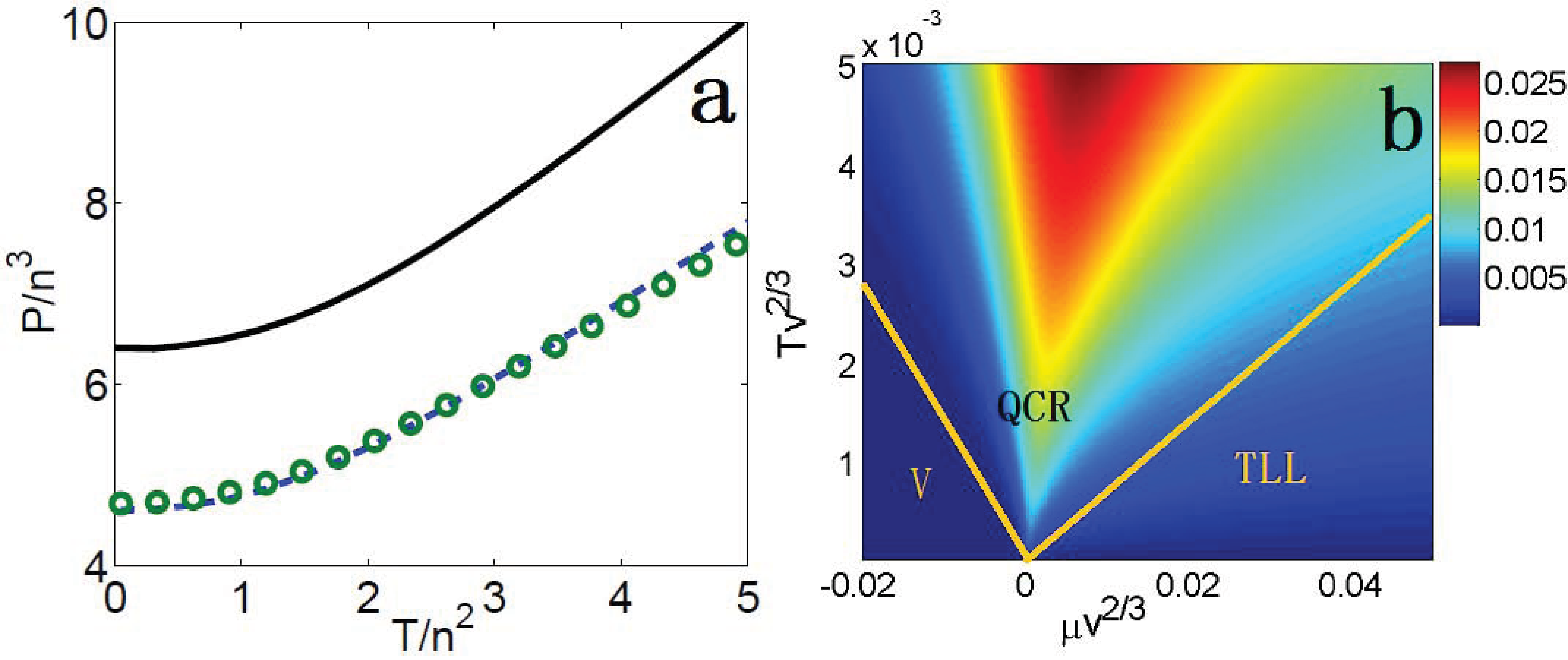}
\caption{Equation of state  of the TG gas ($c_0\rightarrow \infty$) is plotted  by numerically solving equation (\ref{TBA}). Left panel: pressure vs temperature at fixed density for $\tilde{v}=0$ (solid line),  $\tilde{v}=0.003$ (dash line).  The open circles shows the result from analytical result  (\ref{Pplot}). Right panel: entropy in the $T-\mu$ plane.  Where $V$ denotes the vacuum and $QCR$ stands  quantum critical regime. The  right solid line indicates a crossover temperature at which the linear-temperature-dependent entropy  breaks down.  \label{QCP}}
\end{figure}

Furthermore,  we may accurately calculated the thermodynamics of the model from   (\ref{TBA}) in analytic fashion using the polylog function.  For capturing quantum criticality of finite range $v$, we consider the Tonks limit $c_0\to \infty$.
 Following the approach  \cite{polylog},  we  obtain an analytic form of the equation of state
\begin{eqnarray}
P&\!=\!&-\frac{T^{\frac{3}{2}}}{2\sqrt{\pi}}Li_{\frac{3}{2}}\!\big(\!-\!e^{\frac{\mu}{T}+A}\!\big)\Big[1-
\frac{3vT^{\frac{3}{2}}}{2\sqrt{\pi}}Li_{\frac{3}{2}}\big(\!-\!e^{\frac{\mu}{T}+A}\!\big)\Big]\nonumber\\
A&\!=\!&-\frac{3vT^{\frac{3}{2}}}{2\sqrt{\pi}}Li_{\frac{5}{2}}\!\big(\!-\!e^{\frac{\mu}{T}}\!\big)
\!\Big[1 \!-\!\frac{6vT^{\frac{3}{2}}}{\sqrt{\pi}}Li_{\frac{3}{2}}\big(\!-\!e^{\frac{\mu}{T}}\!\big)\Big]\label{Pplot}
\end{eqnarray}
which  is  valid for  finite $v$. This analytical equation of state  is very convenient for experimentalists to evaluate various thermodynamical quantities without involving numerical calculation of  the integral equation (\ref{TBA}). From the figure $(a)$ in Fig. \ref{QCP}, we see that a good  agreement  between analytical result (\ref{Pplot}) and the numerical  result  from  the TBA  equation (\ref{TBA}). The significance of this result is that   the equation of state (\ref{Pplot})  allows the exploration of TLL  physics and quantum criticality in  a highly precise manner.

At low temperatures, from  (\ref{Pplot}) we  further find the universal field theory  of TTL, i.e.,  up to $v^2$,
\begin{eqnarray}
P(\mu,T)&=&P_0+\beta(\mu)\frac{\pi T^2}{6}+O(T^4)\label{Ptot}\\
\beta(\mu)&=&\frac{1}{2\sqrt{\mu}}\Big[1+\frac{48}{5\pi}v\mu^{3/2}+\frac{2856}{25\pi^2}\big(v\mu^{3/2} \big)^{2} \Big]\nonumber
\end{eqnarray}
where pressure  $P_0=\frac{2\mu^{3/2}}{3\pi}\left[1+ \frac{16}{5\pi}v\mu^{\frac{3}{2}}+\frac{616}{25\pi^2}(v \mu^{\frac{3}{2}})^2\right]$.
For  fixed density $n$, the free energy can be written as
\begin{eqnarray}
F(n,T)=E_0-\beta(n)\frac{\pi T^2}{6}+O(T^4)\label{free_energy}
\end{eqnarray}
where $E_0$ is the ground state energy and $\beta(n)$ is given as:
\begin{eqnarray}
\beta(n)=\frac{1}{2n\pi} \Big(1 + 16\pi^2 \tilde{v} - \frac{288}{5}\pi^4 \tilde{v}^2\Big).
\end{eqnarray}
Thus  we have obtained the Gaussian  conformal field theory with  the sound velocity  $\nu_c=1/\beta(n)$ and central charge $C=1$.   We further prove that this  sound velocity does coincide with the calculation from the excitation spectrum (\ref{vc1}).

\section{Quantum Criticality}
By analyzing the TBA  equation (\ref{TBA}) in zero temperature limit,  we see that the quantum phase transition from vacuum to TLL  occurs  as the chemical potential  is  varied through the critical point $\mu_c=0$. Near the  quantum critical point, this  many-body system is expected to show universal scaling behavior in the thermodynamic quantities due to the collective nature of many-body effect.  Taking  $c_0\to  \infty$, the quantum critical phenomena of the model reveal subtle dependence of finite range parameter  $v$. The universal scaling functions  of density and compressibility $\kappa$ near this critical point can be obtained from the equation  of state  (\ref{Pplot}), i.e.
\begin{eqnarray}
n(\tilde{\mu},\tilde{T})-n_0(\tilde{\mu},\tilde{T})&=&\tilde{T}^{\frac{d}{z}+1-\frac{1}{vz}}\mathcal{F}\Big(\frac{\tilde{\mu}-\tilde{\mu}_c}{\tilde{T}^{\frac{1}{vz}}} \Big)\\
\kappa(\tilde{\mu},\tilde{T})-\kappa_0(\tilde{\mu},\tilde{T})&=&\tilde{T}^{\frac{d}{z}+1-\frac{2}{vz}}\mathcal{Q}\Big(\frac{\tilde{\mu}-\tilde{\mu}_c}{\tilde{T}^{\frac{1}{vz}}} \Big)
\end{eqnarray}
where $\mathcal{F}(x)=-Li_{\frac{1}{2}}(-e^x)/(2v^{1/3}\sqrt{\pi})$, $\mathcal{Q}(x)=-\frac{v^{1/3}}{2\sqrt{\pi}}Li_{-\frac{1}{2}}(-e^x)$, $\tilde{T}=Tv^{2/3},~\tilde{\mu}=\mu v^{2/3}$ with the background $n_0=\kappa_0=0$ read off the critical exponents $z=2,\, \nu =1$.  This scaling form looks  similar to the one of the  Lieb-Liniger model \cite{polylog}.  However, in contrast to the scaling behavior of   the  hardcore Bose gas and the  Lieb-Liniger model \cite{Zhou-Ho,polylog,polylog2,polylog3,polylog4} in the wide CIR, here  the scalings  are  rescaled by the effective finite range parameter $v$.
Fig. \ref{QCP}b  shows the universal critical behavior  of entropy density  for  the $c_0\rightarrow \infty$ limit.  The well pronounced fan-shape like phase diagram of  criticality indicates  a universal crossover temperature line that separates  the relativistic TLL from  the non-relativistic free fermion theory.

\begin{figure}[t]
\includegraphics[height=1.4in, width=3.3in]
{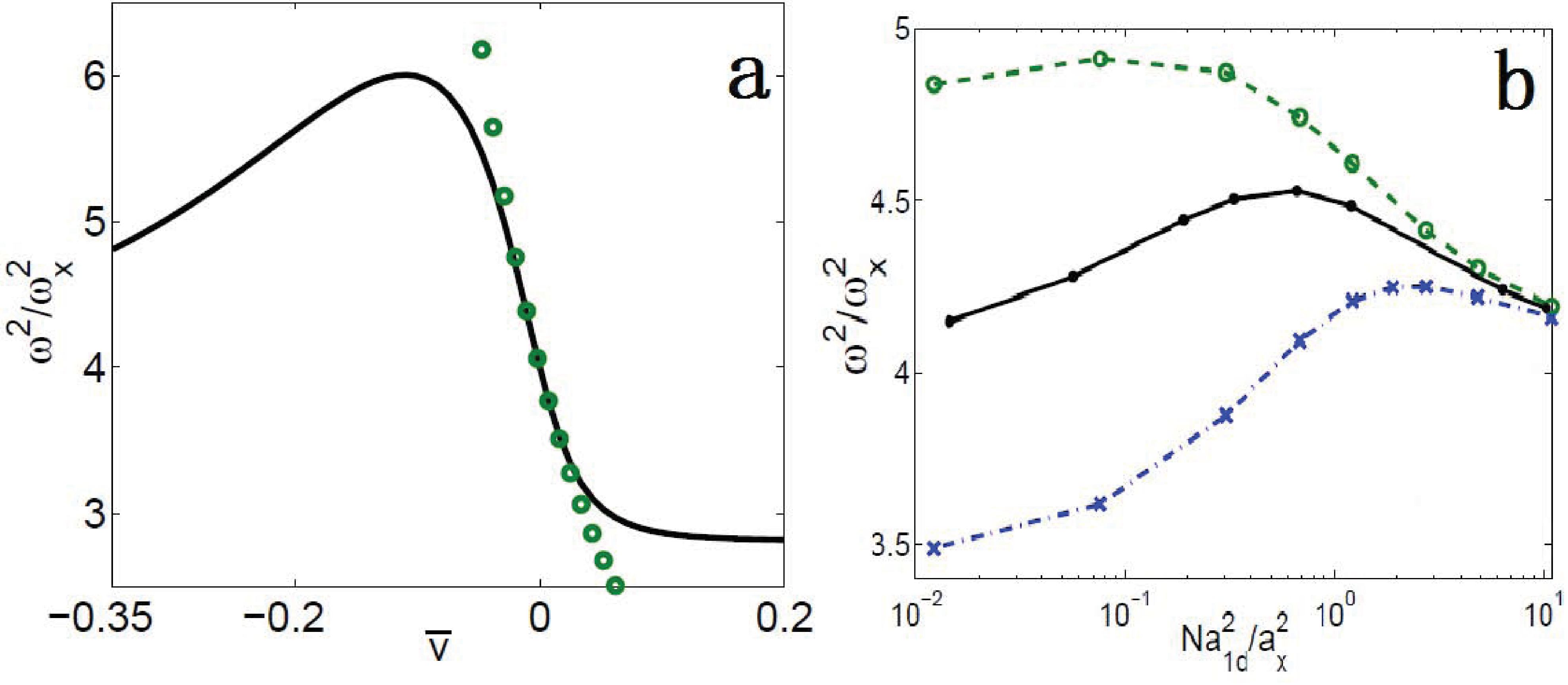}
\caption{Breathing mode frequency across a narrow CIR. Left panel: $\omega$ as a function of $\bar{v}$ in the Tonk limit $c_0\rightarrow\infty$ where solid line is from numerical calculation while open circles refer to small $v$ expansion. Right panel: $\omega$ as a function of zero energy scattering strength in the super-TG regime where solid, dash-dotted, dashed lines refer to $\bar{v}=0,~0.021,~-0.021$.  \label{Breathmode}}
\end{figure}

\section{Experimental signatures}
The oscillation of fundamental modes  in a shallow harmonic trap  (often call breathing mode) is  of particular interest in realistic  experiments. The breathing mode frequency is sensitive to various interaction regimes. Therefore,  it  can be used to precisely determine  different  quantum phases  in current experiments \cite{Liming,liming2}.  Despite of the fact that  for a narrow resonance one always has negative effective range such that $v>0$,  it  is also possible to create a  large positive effective range  $v<0$
 like the case of dipolar interaction induced resonance \cite{DIIR}.  We will show that the finite effective range $v>0$ or $v<0$  have  dramatic influences on the breathing mode frequency $\omega$ across a narrow CIR.

Suppose the 1D system is confined in a shallow harmonic trap $V_{ext}(x)=\frac{1}{2}m\omega_x^2x^2$ along axial direction. The breathing mode frequency $\omega$ from sum rule is given by  $\omega^2/\omega_x^2=-2\langle x^2\rangle/(\partial\langle x^2\rangle/\partial\omega_x^2)$,
where $\langle x^2\rangle=\int x^2\rho(x)dx/N$ and $\rho(x)$ is the density distribution along the trap.  Here $\rho(x)$ can be obtained from the BA equation (\ref{rhokground})  within the  local density approximation. Following the convention used in experiment \cite{superTonk}, we prefer the quantity $A^2=(a_{1d}/\ell_x)^2$ to character the zero energy interaction strength and  $\bar{v}=v/\ell_x^3$ to character the effective range effect in the trap.  In these equations, we have defined $\ell_x=a_x/\sqrt{N}~,a_{1d}=-2/c_0$ and $a_x=\sqrt{1/(m\omega_x)}$.  In Fig .\ref{Breathmode}b, we plot  the ratio of $\omega^2/\omega^2_x$ as a function of $A^2$ in the super-TG regime with $|c_0|\gg 1 $ at zero temperature. One can see that both the value and position of the peak change dramatically for the values of $\bar{v} \simeq 0, \pm 0.02$. This signature  is capable of  being measured  in  sufficiently narrow resonances.  We  also note that although for a given resonance the effective range $r_0$ is fixed,  one can tune the value of $\bar{v}$ in a wide range  by changing the confining length $a_x$. Where $\bar{v}\propto1/a_x^3$.

Moreover, in the ``Tonk limit" $c_0\to  \infty$,  by taking small  $v$ expansions  with the  equation (\ref{Ptot}) at zero temperature,  we obtain $\omega^2/\omega_x^2 \approx 4/(1+a\bar{v}+b\bar{v}^2)$, where $a=2048\sqrt{2}/(35\pi^2)$ and $b=(17827425\pi^2-153092096)/(91875\pi^4)$. We compare this asymptotic expression   with numerical result in Fig .\ref{Breathmode} and see  an  excellent  agreement  between the two results for small values of  $\bar{v}$. We  find  a  peak structure on the negative side of $v$ around $\bar{v}\simeq -0.1$.  This signature indicates  that in the $c_0\to \infty$ limit,  a negative $v$  gives an effective  super-TG  gas like phase at the ground state.

Before concluding, we would like to compare our work to another related paper by Gurarie \cite{Gurarie}. In Ref. \cite{Gurarie}, the author also considered this many-body problem of  a one dimensional boson gas across a narrow resonance.  In Ref. \cite{Gurarie}, the author started with a pure one dimensional two-channel mode and identified some interesting phases in different parameter regimes.  It seems that their analysis does not further indicate whether  these regimes are accessible or not in current cold atom experiments. Here we start with a two-body amplitude which is obtained from a realistic two-body calculation of narrow CIR. Using Bethe ansatz solution, we have shown that  the underlying physics is likely to be  accessible in current CIR experiments. Also, under particular  choices of parameters,  which are  based on realistic experimental conditions, we have obtained  interesting results for  the thermodynamic properties and quantum critical behaviors.

\section{Summary}
In conclusion, using exact  BA solution we have studied  significantly different strong interaction effects in many-body properties  of the  quasi-1D Boson gas across  the narrow CIR. The universal TLL physics, equation of state,  quantum criticality and high temperature thermodynamics have been obtained in terms of  the effective finite range parameter $v$. It turns out that this parameter essentially changes the quantum statistics of the TG gas from free Fermi statistics to  mutual statistics or to the more exclusive super TG gas.   Measuring the breathing mode enables us  to capture these striking  strong interaction  effects in current cold atom experiments.

\acknowledgments
We specially thank Hui Zhai for bringing some basic ideas and very helpful discussions. This work is supported by NSFC under Grant No. 11104157, No. 11004118 and No. 11174176, and NKBRSFC under Grant No. 2011CB921500. XWG is  partially supported by the ARC and the National Basic Research Program of China under Grant No. 2012CB922101.

\end{document}